\documentclass[aps,prb,twocolumn,showpacs]{revtex4}

\usepackage{graphicx}
\usepackage{amsmath} 

\begin{document}


\title{Single polaron properties of the breathing-mode Hamiltonian}
\author{Bayo Lau} \author{Mona Berciu} \author{George A. Sawatzky}
\affiliation{Department of Physics and Astronomy, University of
British Columbia, Vancouver, BC, V6T 1Z1} \date{\today}

\begin{abstract}
We investigate numerically various properties of the one-dimensional (1D)
breathing-mode polaron. We  use an extension of a variational
scheme to compute the energies and wave-functions of the two
lowest-energy eigenstates for any momentum, as well as a scheme to
compute directly the polaron's Green's function. We contrast these results with
results for the 1D Holstein polaron. In particular, we find that the
crossover from a
large to a small polaron is significantly sharper. Unlike
for the Holstein model, at moderate and large couplings the
breathing-mode polaron dispersion has non-monotonic dependence on the
polaron momentum $k$. Neither of these aspects is revealed by a
previous  study based on the self-consistent Born approximation.
\end{abstract}

\pacs{71.38.-k, 72.10.Di, 63.20.Kr}

\maketitle

\section{\label{sec:intro} Introduction}

In a solid-state system, the interaction between a charge carrier
and phonons (quantized lattice vibrations) leads to the formation of
polarons. This mechanism is a key ingredient in the physics of the
manganites\cite{manganite} and, possibly, of the
cuprates.\cite{cuprate}$^,$\cite{TomD} There are various model
Hamiltonians describing the coupling of the particle and bosonic
degrees of freedom. The asymptotic limits of weak or strong coupling
can be investigated analytically using perturbation theory, however
the intermediate-coupling regime generally requires numerical
simulations. Recently, investigations of basic model Hamiltonians
have progressed rapidly thanks to the development of efficient
analytical and computational tools, and we are now able to begin to
study more and more realistic models.

The simplest electron-phonon coupling is described by the Holstein
Hamiltonian.\cite{Holstein} It is essentially the tight-binding model
with an on-site energy proportional to the lattice displacement
$X_i=\frac{1}{\sqrt{2M\Omega}}(b_i^{\dag}+b_i)$:
\begin{equation}
\label{holstein}
H_H =-t \sum_{<ij>} \left(c_i^{\dag}c_j+
c_j^{\dag}c_i\right)+\Omega\sum_{i}b_i^{\dag}b_i + g \sum_{i}^{} n_i X_i
\end{equation}
Here $c_i$ is the annihilation operator for an electron at site $i$
(since we only consider the single electron case, the spin is
irrelevant and we drop its index in the following. We also set
$\hbar =1$).  $t$ is the hopping matrix, $n_i = c_i^\dag c_i$. For
the Einstein phonons, $b_i$ is the annihilation operator at site
$i$, $\Omega$ is the frequency, $M$ is the atomic mass, and $g$ is
the electron-phonon coupling strength. The model has been widely
studied numerically by Monte-Carlo
calculations,\cite{mc1}$^-$\cite{mc12} variational
methods,\cite{bonca}$^-$\cite{v18} and exact
diagonalization.\cite{md1}$^-$\cite{marsiglio} Analytic
approximations have also progressed over the years.\cite{a1,MA,a2,a3}

For some materials, a more appropriate model is provided by the
breathing-mode Hamiltonian. For example, consider a half-filled 2D
copper-oxygen plane of a parent cuprate compound. Injection of an
additional hole should fill an oxygen $2p$ orbital. Due to
hybridization between oxygen $2p$ and copper $3d_{x^2-y^2}$
orbitals, the hole resides in fact in a so-called Zhang-Rice singlet
(ZRS) with a binding energy proportional to $-8t^2_{dp}$, where
$t_{dp}$ is the hopping between neighboring O and Cu orbitals. The
dynamics of the ZRS can be described by an effective one-band model
with orbitals centered on the copper sub-lattice.\cite{ZR} If
lattice vibrations are considered, the motion of the lighter oxygen
ions, which live on the bonds connecting Cu sites, are the most
relevant. The hopping integral $t_{dp}$ and charge-transfer gap
between Cu and O orbitals are now modulated as the oxygen moves
closer or further from its neighboring Cu atom. Both the on-site
energy and hopping integral are modulated in the effective one-band
model, but the former has been shown to be
dominant\cite{br1}$^,$\cite{br2}. The breathing-mode Hamiltonian
describes the physics of the linear modulation of on-site energy.

While this breathing-mode model is motivated as a 2D model, in this
work we investigate numerically only its 1D version, relevant,  {\em
  e.g.}  for CuO chains. In 1D we can investigate accurately and
efficiently not only ground-state (GS) properties, but also some excited
state properties.
For the Holstein model, it was found that
polaron properties are qualitatively similar in different
dimensions,\cite{v6,a2} but with a sharper
large-to-small polaron crossover in higher dimensions. We will show
that a sharp crossover is already present in
the 1D model, and we expect less dimensionality effects in the
breathing-mode Hamiltonian.

The 1D breathing-mode Hamiltonian that we investigate here is
 described by:
\begin{eqnarray}
&& H_B = -t \sum_{i} \left(c_i^{\dag}c_{i+1}+ c_{i+1}^{\dag}c_i
\right)+ \Omega\sum_{i}b_{i+{1\over 2}}^{\dag}b_{i+{1\over
2}}\nonumber\\
&&+\frac{g}{\sqrt{2M\Omega}}\sum_{i}n_i\left(b_{i+{1\over
2}}^{\dag}+b_{i+{1\over
2}}-b_{i-{1\over2}}^{\dag}-b_{i-{1\over2}}\right)
\label{breathing}
\end{eqnarray}
The notation is the same as before, except that now the phonons live
on an interlaced lattice. The difference between the two models is
more apparent in momentum space. The Holstein model has constant
coupling to all phonon modes
\begin{equation}
\label{eq1} V_{H} =
\frac{g}{\sqrt{N}\sqrt{2M\Omega}}\sum_{kq}c^{\dag}_{k-q}c_k
\left(b^{\dag}_q+ b_{-q} \right),\nonumber
\end{equation}
whereas the breathing-mode model has a coupling strength that
increases monotonically with increasing phonon momentum
\begin{equation}
\label{eq2} V_{B} =
\frac{2ig}{\sqrt{N}\sqrt{2M\Omega}}\sum_{kq}\sin{\frac{q}{2}}c^{\dag}_{k-q}c_k
\left(b^{\dag}_q + b_{-q} \right).\nonumber
\end{equation}
Here $N$ is the number of lattice sites, and becomes infinite in the
thermodynamic limit. The momenta $k,q$ are restricted to the first
Brillouin zone $(-\pi, \pi]$ (we take the lattice constant $a=1$).

While numerical and analytical studies of the Holstein polaron
abound, there is much less known about models with $g(q)$ coupling.
In particular, there is  no detailed numerical study of the 1D
breathing-mode model, apart from an exact diagonalization of a simplified
t-J-breathing-mode model in restricted basis \cite{bed}, and an
investigation based on the self-consistent Born approximation
\cite{scba}, which is known to become inaccurate for intermediate
and strong couplings. In this work, we study numerically various
low-energy properties and the spectral function of the single
polaron in the 1D breathing-mode Hamiltonian. The results are
compared with the relevant results for the single Holstein 1D
polaron, allowing us to contrast the behavior of the polarons in the
two models. The article is organized as follows: in Sec.
\ref{sec:Methology} we review relevant asymptotic results, and
describe the numerical methods we use to calculate low-energy
properties and the spectral functions for both models. In Sec.
\ref{sec:results} we present our results, and in Sec. \ref{sec:conc}
we present our conclusions.

\section{\label{sec:Methology} Methodology}
\subsection{\label{sec:perturbation}Strong-coupling perturbation results}

Perturbational results for the strong-coupling limit $g\gg t$
provide a good intuitive picture of the problem even for the
intermediate coupling regime. In the absence of hopping, $t=0$, both
Hamiltonians can be diagonalized by the Lang-Firsov
transformation\cite{Mahan}
\begin{equation}
\tilde{O}=e^{S}O e^{-S}.\label{eq:LFT}
\end{equation}
Using
\begin{equation}
S_H=\frac{g}{\Omega\sqrt{2M\Omega}}\sum_i
n_i(b_i^{\dag}-b_i),\nonumber
\end{equation}
and
\begin{equation}
S_B=\frac{g}{\Omega\sqrt{2M\Omega}} \sum_i n_i(-b_{i-{1\over
2}}^{\dag}+b_{i-{1\over 2}} +b_{i+{1\over 2}}^{\dag}-b_{i+{1\over
2}})\nonumber
\end{equation}
respectively, the diagonal forms of the Hamiltonians are, in terms of
the original (undressed) operators:
\begin{equation}
\tilde{H}_H=\tilde{T}_H+\Omega\sum_i b^{\dag}_{i+{1\over
2}}b_{i+{1\over2}} - \frac{g^2}{2M\Omega^2}\sum_i n_i^2\label{eq:LFH}
\end{equation}
\begin{equation}
\tilde{H}_B=\tilde{T}_B+\Omega\sum_i b^{\dag}_{i+{1\over 2}}
b_{i+{1\over 2}} - \frac{2g^2}{2M\Omega^2}\sum_i n_i(n_i -
n_{i+1})\label{eq:LFB}
\end{equation}
where the kinetic energies are:
\begin{equation}
{\tilde T}_H=-te^{-\frac{g^2}{2M\Omega^3}}\sum_i c_{i+1}^{\dag}c_i
e^{\frac{g(-b_{i+1}^{\dag}+b_i^{\dag})}{\Omega\sqrt{2M\Omega}}}
e^{-\frac{g(-b_{i+1}+b_i)}{\Omega\sqrt{2M\Omega}}} +h.c.\nonumber
\end{equation}
\begin{eqnarray}
\tilde{T}_B=&&-te^{-3\frac{g^2}{2M\Omega^3}}\sum_i c_{i+1}^{\dag}c_i
e^{\frac{g}{\Omega\sqrt{2M\Omega}}(b_{i+{3\over
2}}^{\dag}-2b_{i+{1\over 2}}^{\dag}+b_{i-{1\over 2}}^{\dag})}\times
\nonumber\\ &&e^{-\frac{g}{\Omega\sqrt{2M\Omega}}(b_{i+{1\over
2}}-2b_{i-{1\over 2}}+b_{i-{3\over2}})}+ h.c.\nonumber
\end{eqnarray}
For a $d$-dimensional lattice, $\tilde{T}_B$ is modified by i) extra
creation and annihilation operators of phonons in the direction
transverse to hopping, and ii) change of the -3 factor in the exponent
to $-(z+1),
z=2d$. The third term in Eq.~(\ref{eq:LFH}) and in Eq.~(\ref{eq:LFB})
signifies that the mere presence of an electron would induce a lattice
deformation, leading to the formation of a  polaron to lower the system's
energy. For a single polaron, the  lattice deformation energy
is proportional to the number of nearest phonon sites (one for the
Holstein model and $z$ for the breathing-mode model). For  $t=0$, the
ground-state
energy is degenerate over momentum-space:
\begin{eqnarray}
E^{(0)}_{H}(k)=-\frac{g^2}{2M\Omega^2}\nonumber\\
E^{(0)}_{B}(k)=-z\frac{g^2}{2M\Omega^2}\nonumber
\end{eqnarray}
Each model has three energy scales, therefore the parameter space can be
characterized by two dimensionless ratios. It is natural to define the
dimensionless (effective) coupling  as the ratio of the lattice
deformation energy to the free-electron ground-state energy $-zt$:
\begin{eqnarray}
\lambda_{H}=\frac{g^2}{2M\Omega^2}\frac{1}{zt}\label{eq:lambda_h}\\
\lambda_{B}=\frac{g^2}{2M\Omega^2}\frac{1}{t}\label{eq:lambda_b},
\end{eqnarray}
where $z$ is also the number of nearest neighbors in the electron
sublattice. It should be noted that since $\Omega\sim1/\sqrt{M}$, the
$\lambda$'s do not depend on the ion mass, $M$. $\lambda$ has been shown
to be a good parameter to describe the large-to-small polaron
crossover in the Holstein model. It will be shown in later sections
that the definition also works well for the breathing-model model.
The other parameter is the adiabatic ratio which appears naturally
from the perturbation in $t$
\begin{equation}
\alpha=\frac{zt}{\Omega}.
\end{equation}

Using standard perturbation theory, \cite{marsiglio} the first-order
corrections to the energy of the lowest state of momentum $k$ are:
\begin{eqnarray}
E^{(1)}_{H}(k)=-2te^{-\alpha\lambda_{H}}\cos(k),\label{eq:e1h}\\
E^{(1)}_{B}(k)=-2te^{-\frac{3}{2}\alpha\lambda_{B}}\cos(k),\label{eq:e1b}
\end{eqnarray}
showing that the polaron bandwidth is exponentially suppressed in the strong
coupling limit. As is well known, this is due to the many-phonons
clouds created on the electron site (Holstein) or on the two phonon
sites bracketing the electron site (breathing-mode model). As the
polaron moves from one site to the next, the overlaps between the
corresponding clouds become vanishingly small and therefore $t_{\rm
eff} \rightarrow 0$. To first order in $t$, the suppression is stronger
for the breathing-mode model simply because the overlap integral
involves phonon clouds on 3 sites instead of just 2, as in the case
for Holstein. The second-order corrections are:
\begin{eqnarray}
E^{(2)}_{H}(k)=-2\frac{t^2}{\Omega}e^{-2\alpha\lambda_{H}}
f_H(k,\alpha,\lambda_H), \label{eq:e2h}\nonumber\\
E^{(2)}_{B}(k)=-2\frac{t^2}{\Omega}e^{-3\alpha\lambda_{B}}
f_B(k,\alpha,\lambda_B).\nonumber
\end{eqnarray}
The functions $f$ can be written in the form
\begin{equation}
f_{H,B}=A_{H,B}+B_{H,B}\cos(2k)\nonumber
\end{equation}
with
\begin{eqnarray}
A_H&=&Ei(2\alpha\lambda_H)-\ln(2\alpha\lambda_H)-\gamma\nonumber\\
B_H&=&Ei(\alpha\lambda_H)-\ln(\alpha\lambda_H)-\gamma\nonumber
\end{eqnarray}
and
\begin{eqnarray}
A_B&=&Ei(3\alpha\lambda_B)-\ln(3\alpha\lambda_B)-\gamma\nonumber\\
B_B&=&Ei(2\alpha\lambda_B)-2Ei(\alpha\lambda_B)+
\ln(\frac{\alpha\lambda_B}{2})+\gamma.\nonumber
\end{eqnarray}
Here, $\gamma$ is the Euler-Mascheroni constant and $Ei(x)$ is the
exponential integral with the series expansion
\begin{equation}
Ei(x)=\gamma+\ln(x)+\sum_{n=1}^{\infty}\frac{x^n}{n!n}.\nonumber
\end{equation}

The result can be further simplified in the limit
$\alpha\lambda_H,\alpha\lambda_B \gg1$ using $\lim_{x->\infty}
\sum_{n=1}^{\infty}\frac{x^n}{n!n} \sim \frac{e^x}{x}$. This leads to
the simplified expressions:
\begin{eqnarray}
E^{(2)}_{H}(k)\sim-2\frac{t^2}{\alpha\lambda_H}\left[\frac{1}{2} +
e^{-\alpha\lambda_{H}}\cos(2k)\right]\label{eq:e2hn}\\
E^{(2)}_{B}(k)\sim-2\frac{t^2}{\alpha\lambda_B}\left[\frac{1}{3} +
\frac{e^{-\alpha\lambda_{B}}}{2}\cos(2k)\right]\label{eq:e2bn}
\end{eqnarray}

Thus, the breathing-mode model's
ground state energy is slightly higher for any finite $t$. For the
Holstein model, the dispersion is monotonic, since the second order
$\cos(2k)$ contribution is suppressed more strongly than the first
order $\cos(ka)$ contribution. However, a quick comparison between
Eq.~(\ref{eq:e1b}) and (\ref{eq:e2bn}) shows that  in
the breathing-mode model, the second order
$\cos(2k)$ contribution becomes dominant at large enough coupling. As
a result, at strong couplings we expect
the breathing-mode polaron energy to exhibit a maximum at a finite $k
< \pi$, and then to fold back down.

\subsection{\label{sec:matcomp}Matrix computation}

The computation method we use is a direct generalization of the method
introduced by Bonca {\em et al.} for the Holstein model, in
Ref. \onlinecite{bonca}. This approach requires sparse matrix
computations to solve the problem. Although expensive cpu and memory
resources are required for this type of method, it gives us a
systematic way to compute excited state properties, which would be
more difficult to achieve using other numerical methods.

The idea is to use a suitable basis in which to represent the
Hamiltonian as a sparse matrix. For the Holstein model, this basis
contains states of the general form\cite{bonca}
\begin{equation}
|S,K\rangle = \frac{1}{\sqrt{N}}\sum_j
e^{iKj}c^{\dag}_j\prod_{m\epsilon\{S\}}\frac{b_{j+m}^{\dag
    n_m}}{\sqrt{n_m!}}|0\rangle,\label{eq:basis}
\end{equation}
where $K$ is the total momentum and S denotes a particular phonon
configuration, with sets of $n_m$ phonons located at a distance $m$
away from the electron. For the Holstein model $m$ are integers, since
the phonons are located on the same lattice as the electrons.  The
generalization for the breathing-mode is simple: here $m$ are
half-integers, since here the phonons live on the interlaced
sublattice. All states in either basis can be obtained by repeatedly
applying the Hamiltonian to the free electron state which has all
$n_m=0$. The possible matrix elements are $-te^{\pm iK}$, $\Omega n$,
and $\pm \frac{g}{\sqrt{2M\Omega}}\sqrt{n}$, where $n$ are integers
related to the numbers of phonons.

Since the Hilbert space of the problem is infinite, this basis must be
truncated for computation. The original cut-off scheme in Ref.
\onlinecite{bonca} was optimized for computation of ground-state
properties of the Holstein model, by restricting the number of matrix
elements between any included state and the free-electron
state. Getting the higher-energy states is more involved, as it is
evident from Eq.~(\ref{eq:LFT}) that each state $|\phi\rangle_{LF}$ in
the Lang-Firsov basis correspond to a state
\begin{equation}
|\phi\rangle_{R}=e^{-S}|\phi\rangle_{LF}\label{eq:transform}
\end{equation}
in real space. With our choice for the $S$ operators, this reverse
transformation induces a phonon coherent state structure at the
electron site (Holstein), respectively the two bracketing phonon
sites (breathing-mode). The phonon statistics of the coherent state
obeys the Poisson distribution. In the anti-adiabatic regime
$(zt>\Omega)$, the splitting due to the hopping (off-diagonal
hopping matrix elements) is significant compared to the diagonal
matrix elements proportional to $\Omega$. The underlying Lang-Firsov
structure needs to be modeled by the hopping of the electron away
from the coherent state structure created by the $e^{-S}$ operator.
To capture these characteristics, the basis is divided into
subspaces with fixed numbers of phonons. Each subspace is enlarged by
the addition of states with phonons further and further away from
the electron site (increase of maximum value of $m$ in
Eq.~(\ref{eq:basis})), until convergence is reached. This procedure
allows for efficient generation of all basis states required to
model the higher-energy states.

Matrices of dimension up to $10^6-10^7$ were needed to compute the two
lowest-energy states accurately. These two states were calculated
numerically using the Lanczos method with QR shift,
\cite{ARPACK,note1} which works efficiently for the low energy bound
states.

The number of bytes required to store a $n\times n$ sparse complex
matrix is roughly $n( (16+4)m+4)$, where $m$ is the number of matrix
elements per row. The number of bytes required to store a $n$-vector
is $16n$. Therefore, an ordinary workstation can deal with
$n\sim10^7$, sufficient for our low-energy states computation. Cluster
parallelization provides decent speed-up up to $n\sim5\times10^7$,
above which communication costs proved to be too high due to the large
matrix bandwidth, even after re-ordering. For the larger $n$ values
used in the Green's function calculation (see below), SMP machines
with high memory-to-cpu ratio were used for efficient computation.

\begin{table}
\caption{\label{tab:cutoff_b} Most-relaxed cut-off condition for the
1D breathing-mode polaron Green's function computation.}
\begin{ruledtabular}
\begin{tabular}{ccc}
Subspace's \# phonons & $|m|_{max}$ & \# States\\ \hline 1-11 & 22.5 -
(\# phonons) & 17053356\\ 12-13 & 10.5& 16871582\\ 14-15 & 9.5 &
28274774\\ 16-17 & 8.5 & 33423071\\ 18-20 & 7.5 & 41757650\\ 21-30 &
5.5 & 42628080\\ 31-40 & 4.5 & 38004428\\ 41-50 & 3.5 & 12857573\\
\end{tabular}
\end{ruledtabular}
\end{table}
Table~\ref{tab:cutoff_b} lists the most-relaxed cut-off condition we
used to calculate the Green's function (see next section) for the 1D
breathing-mode model. Because the queue time is roughly independent
of memory requirements, but is longer than the computation time on
the SMP machines, we relaxed the cut-off condition rather roughly
until convergence was observed. As a result, the size of these
matrices is
 certainly much larger than it has to be.

\subsection{\label{sec:gcomp}Green's function computation}
Computation of higher-energy properties requires much larger
matrices. The memory and fflops needed for such computations are
formidable, especially to extensively investigate the
multi-dimensional parameter space $(\lambda,\alpha,K)$. Furthermore,
one characteristic of the single polaron problem is a continuum of
states starting at one phonon quantum above the $K=0$ groundstate.
Lanczos-type methods typically are problematic in dealing with bands
of eigenvalues with small separation. Therefore, in order to study
higher energy states, we calculate directly the Green's function:\cite{dagotto}
\begin{equation}
G(\omega,k)=\langle k| \frac{1}{\omega-H+i\eta}|k\rangle,
\label{eq:gf}
\end{equation}
where $|k\rangle = c_k^\dag|0\rangle$. This can be written as the
solution of a linear system of equations:
\begin{eqnarray}
G(\omega,k)&=&\langle k|y\rangle\nonumber\\
(\omega-H+i\eta)|y\rangle&=&|k\rangle\nonumber
\end{eqnarray}
One can iteratively tri-diagonalize H by the vanilla Lanczos
process:\cite{demmel}
\begin{equation}
H=QTQ^{\dag}\nonumber
\end{equation}
\begin{equation}
(\omega+i\eta-T)Q^{\dag}|y\rangle=Q^{\dag}|k\rangle\nonumber
\end{equation}
If the RHS is of the form {$[1 0 0...]^T$}, Crammer's rule can be used to
express {$G=\langle k|QQ^{\dag}|y\rangle$} as a continuous fraction in
terms of the matrix elements of the tri-diagonal matrix
{$(\omega+i\eta-T)$}. In particular, this condition is achieved by
picking the initial Lanczos vector to be $|k\rangle$. This method is
efficient because it does not require the complete solution of the
linear system nor of the eigenvalue problem. It is well known that this
type of iterative process suffers from numerical instability, which
leads to the loss of orthogonality in $Q$ and incorrect eigenvalue
multiplicity in $T$. \cite{PRO} We perform the vanilla Lanczos
tri-diagonalization and re-orthogonalize each states in $Q$ against
the starting vector $|k\rangle$ to validate the continuous fraction
expansion. Then, numerical errors may come from the fact that
$T$ may have the wrong eigenvalue multiplicity. However, this will not
affect the
location of poles in the spectral weight, {\em i.e.} the
eigenenergies are accurate.

\section{\label{sec:results} Results}
\subsection{\label{sec:eigenpairs}Low-energy states}

\begin{figure}[t]
\includegraphics[width=0.9\columnwidth]{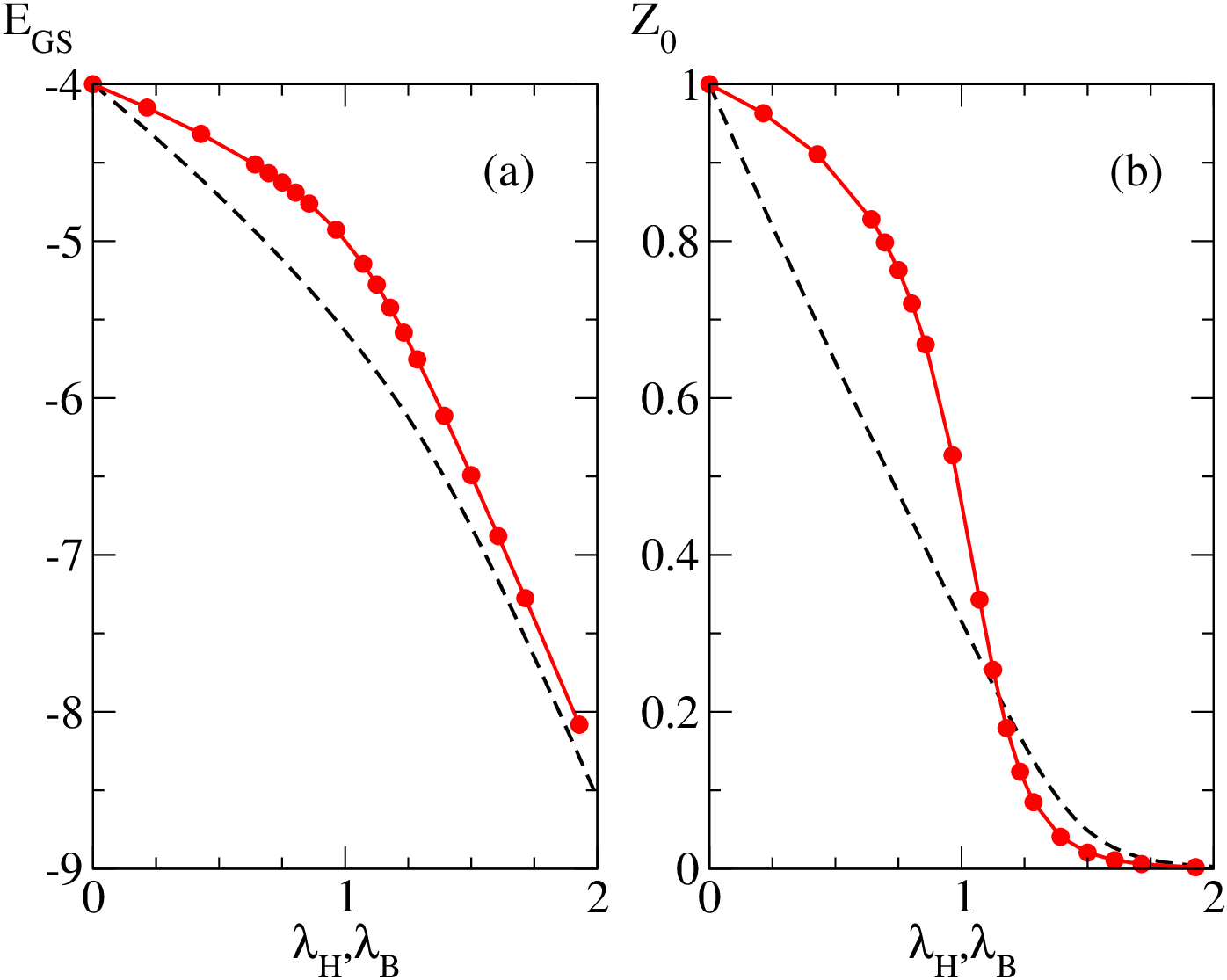}
\caption{\label{fig:gs} (color online) (a) GS energy and (b) GS
  $qp$ weight  as a function of the corresponding
  dimensionless coupling
parameter. The dashed line corresponds to the Holstein model, while
  the breathing-mode results are shown by circles (line is a guide
  to the eye). Parameters are $t=2, \Omega=1$. }
\end{figure}

The ground state energy $E_{GS}$ and quasiparticle ($qp$) weight
$Z_0=|\langle \Phi_{GS}|c^\dag_{k=0}|0\rangle|^2$, where
$|\Phi_{GS}\rangle$ is the ground-state eigenfunction, are shown in
Fig.~\ref{fig:gs} for the 1D breathing-mode and Holstein models. For a
fixed value of $\alpha$, we see the expected crossover from a large
polaron (at weak coupling $\lambda_{B,H}$) to a small polaron (at
strong coupling $\lambda_{B,H}$), signaled by the collapse of the
$qp$ weight.

The ground state energy of both models decreases monotonically with
increasing coupling, but that of the Holstein polaron is lower. This
is in agreement with the second order strong-coupling perturbation
results in Eq.~(\ref{eq:e2hn}) and (\ref{eq:e2bn}). Unlike the rather
gradual decrease in the quasiparticle weight of the 1D Holstein
polaron, the 1D breathing-mode polaron shows a large $Z_0$ at weak
couplings, followed by a much sharper collapse at the crossover near
$\lambda_B \approx 1$. The reason for the enhanced $Z_0$ at weak
couplings is straightforward to understand. Here, the wave-function is
well described by a superposition of the free electron and
electron-plus-one-phonon states. Given the conservation of the total
polaron momentum $K= 0 = k + q$ and the large electron bandwidth $t$,
states with high electron and phonon momenta have high energies and
thus contribute little to the weak-coupling polaron ground-state. On
the other hand, the coupling $g(q)\sim\sin(q/2)$ to the low-energy
states with low electron and phonon momenta is very small for the
breathing-mode model. This explains the slower transfer  of spectral weight
at weak coupling for breathing-mode versus the Holstein
polaron.

\begin{figure}[t]
\includegraphics[width=0.9\columnwidth]{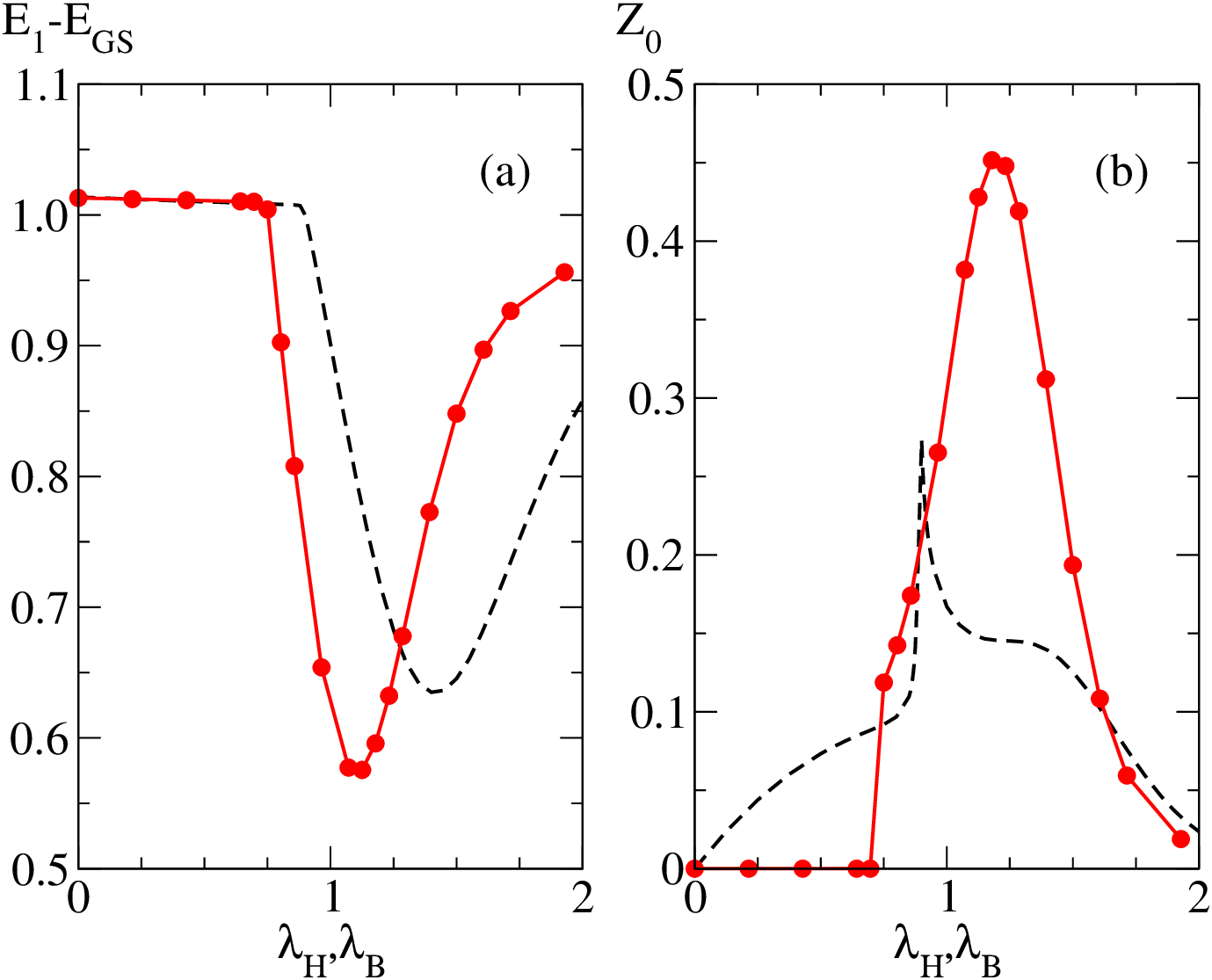}
\caption{\label{fig:1st} (color online) (a) Energy of the first excited
  $K=0$ state, measured from $E_{GS}$, and (b) its
  $qp$ weight. The dashed line corresponds to the Holstein model, while
  the breathing-mode results are shown by circles (line is a guide
  to the eye). Parameters are $t=2, \Omega=1$. }
\end{figure}

The energy (measured from $E_{GS}$) and $qp$ weight of the first
excited $K=0$ state are shown in
Fig.~\ref{fig:1st}. For both
models, at weak-coupling this state is precisely at $\Omega$
above the ground-state energy, at the lower edge of the
polaron+one-phonon continuum. As the coupling  increases
above a critical value, a second bound state gets pushed below the
continuum. This second bound state is absent in SCBA
calculations.\cite{scba} The separation between the two lowest energy
states now first decreases and then increases back towards $\Omega$ as
$\lambda_{H,B} \rightarrow \infty$. This behavior is well-known for
the Holstein polaron.\cite{bonca} The breathing-mode polaron shows the same
qualitative behavior. Note that below the critical coupling, the
computed energy of the first excited state is slightly larger than
$\Omega$. The reason is a systematic error that can be reduced by
increasing the number of one-phonon basis states, in order to better
simulate the delocalized phonon that appears in this state. The
$qp$ weight of the first excited state is zero below the
critical coupling, due to the crossing
between on-site and off-site phonon states.\cite{bonca}

\begin{figure}[b]
\includegraphics[width=1.0\columnwidth]{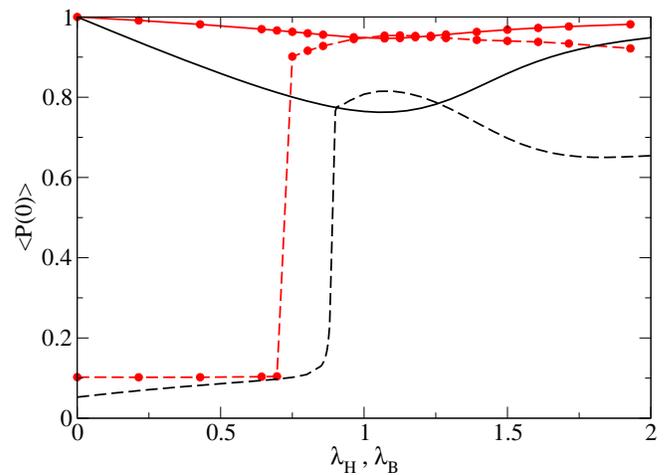}
\caption{\label{fig:os}  (color online) P(0) for $t=2,
\Omega=1$ for the breathing-mode (red symbols) and Holstein (black
line) models, respectively. The solid and dashed lines correspond to
ground state and first excited state, respectively.}
\end{figure}

The nature of the these states is revealed by checking the locality of
the phonon cloud.  We define the projection operator
\begin{equation}
P(K)=\sum_{S_{local}} |S,K\rangle\langle S,K|\nonumber
\end{equation}
where the summation is over all states with $m_H=0$ and $m_B=\pm0.5$
in Eq.~(\ref{eq:basis}). Comparison with Eq.~(\ref{eq:transform})
shows that this operator selects only basis states with phonons only
on the electron site (Holstein)
and only on the two phonon sites bracketing the electron site
(breathing-mode).  Fig.~\ref{fig:os} shows the expectation value of
this operator for the two lowest eigenstates of both models. For both
ground-states $\langle P(0)\rangle \sim 1$, indicating that here most
phonons are neariest to the electron. However, at weak coupling the first
excited state (which is here the band-edge of the polaron + free
phonon continuum) has $\langle P(0)\rangle \rightarrow 0$, precisely
because the free phonon can be anywhere in the system. When the second
bound states form,
$\langle P(0)\rangle$ becomes large,
showing that phonons in these states are primarily localized near the
electron.\cite{bonca} While there appears to be a crossing
between the ground-state and first-excited state values, we emphasize
that $P(K)$ measures the locality of the phonon cloud, not its
structure.

\begin{table}
\caption{\label{tab:fitgs} $\alpha_{n_-,n_+}$ vs. $n_+, n_-$ for the
ground state}
\begin{ruledtabular}
$t=2$  $\Omega=1$  $g=0$  $\Delta=0$
\begin{tabular}{c|c c c c}
$n_+ n_-$ & $0$ & $1$ & $2$ & $3$ \\
\hline
0 & 1.0000 & 0.0000 & 0.0000 & 0.0000\\
1 & 0.0000 & 0.0000 & 0.0000 & 0.0000\\
2 & 0.0000 & 0.0000 & 0.0000 & 0.0000\\
3 & 0.0000 & 0.0000 & 0.0000 & 0.0000\\
\end{tabular}
$t=2$  $\Omega=1$  $g=1.5$  $\Delta=1.05$
\begin{tabular}{c|c c c c}
$n_+ n_-$ & $0$ & $1$ & $2$ & $3$ \\
\hline
0 & 0.9150 & 0.0584 & 0.1372 & -0.0477\\
1 & -0.0584 & -0.1681 & 0.0520 & -0.0536\\
2 & 0.1372 & -0.0520 & 0.0549 & -0.3300\\
3 & 0.0477 & -0.0536 & 0.0330 & -0.0255\\
\end{tabular}
$t=2$  $\Omega=1$  $g=1.964$  $\Delta=1.964$
\begin{tabular}{c|c c c c}
$n_+ n_-$ & $0$ & $1$ & $2$ & $3$ \\
\hline
0 & 0.9876 & -0.0509 & 0.0082 & -0.0025\\
1 & 0.0509 & -0.0100 & 0.0034 & -0.0020\\
2 & 0.0082 & -0.0034 & 0.0022 & -0.0019\\
3 & 0.0025 & -0.0020 & 0.0019 & -0.0019\\
\end{tabular}
\end{ruledtabular}
\end{table}

For the breathing-mode model, these results  suggest the possibility
to describe them using the
on-site coherent-state structure. That is, a Lang-Firsov state with
$n_-$ and $n_+$ number of phonons excited to the left and right of the
electron site, mapped to real space by Eq.~(\ref{eq:transform}). We
note that we are no longer in the strong coupling regime and the
transformation cannot be determined by $g$ and $\Omega$ alone,
therefore we seek an effective transformation with
\begin{equation}
\tilde{S}(\Delta)=S_B|_{\frac{g}{\Omega}=\Delta}\nonumber
\end{equation}
The computed eigenstates $|\phi\rangle$ are projected into such
structure $\alpha_{n_-,n_+}$ by
\begin{eqnarray}
&&\frac{1}{\sqrt{N}}\sum_l e^{iKl} c_l^\dag\sum_{n_-,n_+=1}^\infty
\alpha_{n_-,n_+} \frac{(b_{l-{1\over 2}}^\dag)^{n_-}}{\sqrt{n_-!}}
\frac{(b_{l+{1\over 2}}^\dag)^{n_+}}{\sqrt{n_+!}}|0\rangle\nonumber\\
&&=e^{\tilde{S}(\Delta)} P(K) |\phi\rangle\nonumber
\end{eqnarray}

\begin{table}
\caption{\label{tab:fit1s} $\alpha_{n_-,n_+}$ vs. $n_+, n_-$ for the
first-excited state}
\begin{ruledtabular}
$t=2$  $\Omega=1$  $g=1.5$  $\Delta=1.05$
\begin{tabular}{c|c c c c}
$n_+ n_-$ & $0$ & $1$ & $2$ & $3$ \\
\hline
0 & 0.0228 & -0.6515 & 0.0152 & 0.1151\\
1 & 0.6515 & 0.0062 & 0.1700 & -0.0473\\
2 & 0.0152 & -0.1700 & 0.0505 & -0.0456\\
3 & 0.1151 & -0.0473 & 0.0456 & -0.0220\\
\end{tabular}
$t=2$  $\Omega=1$  $g=1.964$  $\Delta=1.964$
\begin{tabular}{c|c c c c}
$n_+ n_-$ & $0$ & $1$ & $2$ & $3$ \\
\hline 0 & -0.0859 & -0.6675 & 0.0630 & -0.0164\\
1 & 0.6675 & -0.0864 & 0.0246 & -0.0120\\
2 & 0.0630 & -0.0246 & 0.0134 & -0.0099\\
3 & 0.0164 & -0.0120 & 0.0099 & -0.0089\\
\end{tabular}
\end{ruledtabular}
\end{table}

Table ~\ref{tab:fitgs} and \ref{tab:fit1s} show the results of such
projections for the ground state and for the first excited {\em
bound} state. It is clear that they can be rather well described as
$|GS\rangle \sim e^{-\tilde{S}(t,g,\Omega,K)}
\frac{1}{\sqrt{N}}\sum_l e^{iKl} c_l^\dag |0\rangle $, respectively
$|1\rangle_{\rm bound} \sim e^{-\tilde{S}(t,g,\Omega,K)}
\frac{1}{\sqrt{N}}\sum_l e^{iKl} c_l^\dag (e^{i\theta}b_{l-{1\over
2}}^\dag-e^{-i\theta}b_{l+{1\over 2}}^\dag)|0\rangle$ for some phase
$\theta(t,g,\Omega,K)$ needed to satisfy time-reversal symmetry.
These states no longer have definite parity symmetry like those of
the Holstein model. The symmetry is broken by the anti-symmetric coupling term in the model. If the free-electron component is non-zero for an eigenstate, its components with odd/even number of phonons should have odd/even parity. For increasing momentum, this description is valid as long as the excited  state remains bound, with energy less than $E_{GS}+\Omega$.

\begin{figure}
\includegraphics[width=0.9\columnwidth]{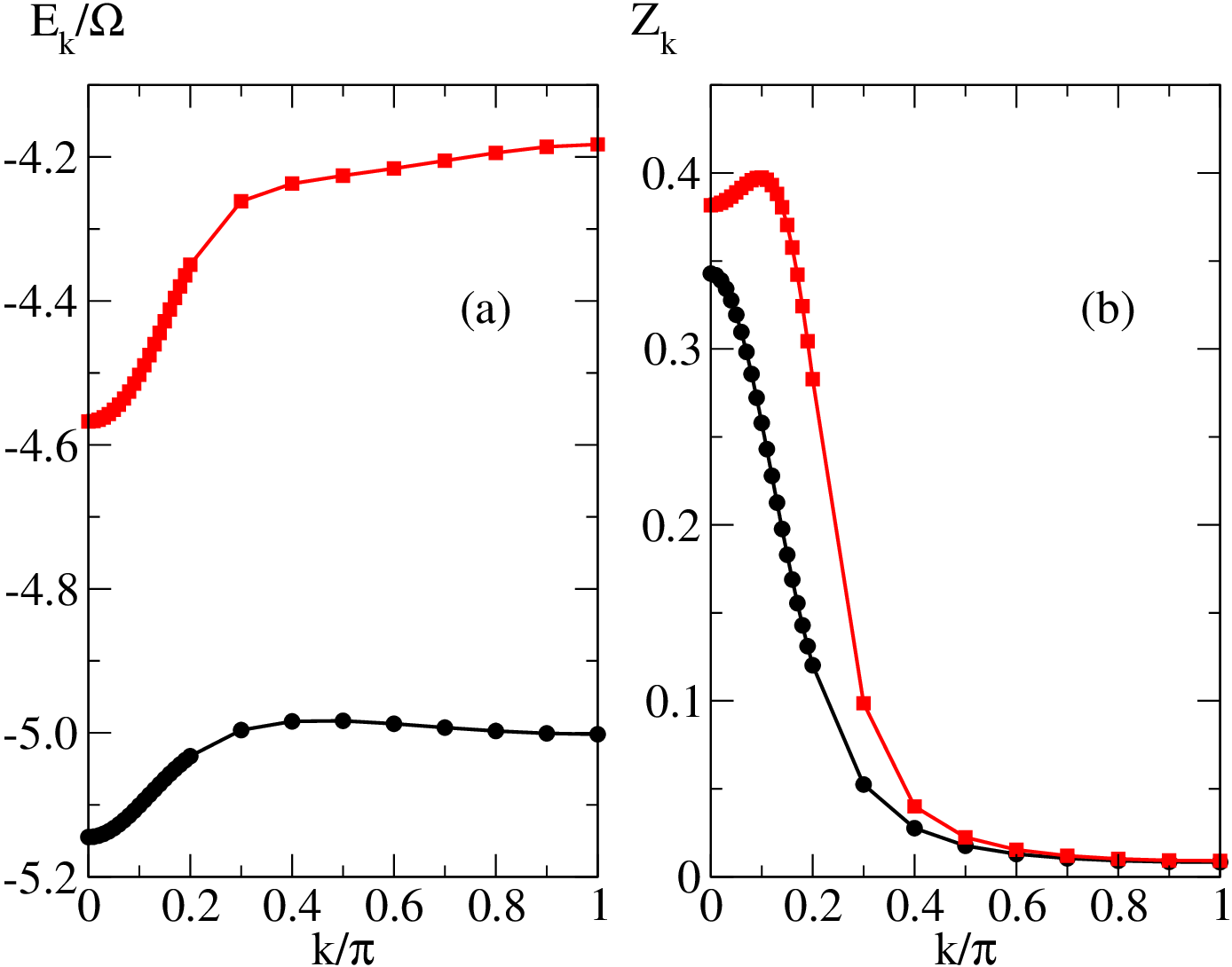}
\caption{\label{fig:disp} (color online) The $k$-dependent (a) energy
  and (b)
  $qp$ weight
of the GS (circles) and first bound state (square) for the 1D breathing mode
  model, for $t=2,
\Omega=1, \lambda_B=1.07$. }
\end{figure}

Fig.~\ref{fig:disp} shows momentum dependent energy and $qp$ weight
for the two lowest eigenstates of the breathing-mode model, for an
intermediate coupling strength $\lambda_B =1.07$. The polaron band has
a maximum at $k \sim0.4\pi$, in qualitative agreement
with the strong-coupling perturbation theory results. This behavior is not
captured by SCBA which is only accurate for low coupling
strength.\cite{scba} For the Holstein polaron, the
polaron dispersion is a
monotonic function of momentum.\cite{bonca} Even though the $qp$
weights remain moderately high at zero momentum, the weights collapse
towards zero with increasing momentum, similar to the well-known
Holstein case. This is due to the fact that at large total
momentum, the significant contribution to the eigenstate comes from
states with at least one or more phonons. The free electron state has
a large energy for large momentum, and contributes very little to the
lowest energy eigenstates, so indeed $Z\rightarrow 0$.

\begin{figure}
\includegraphics[width=0.8\columnwidth]{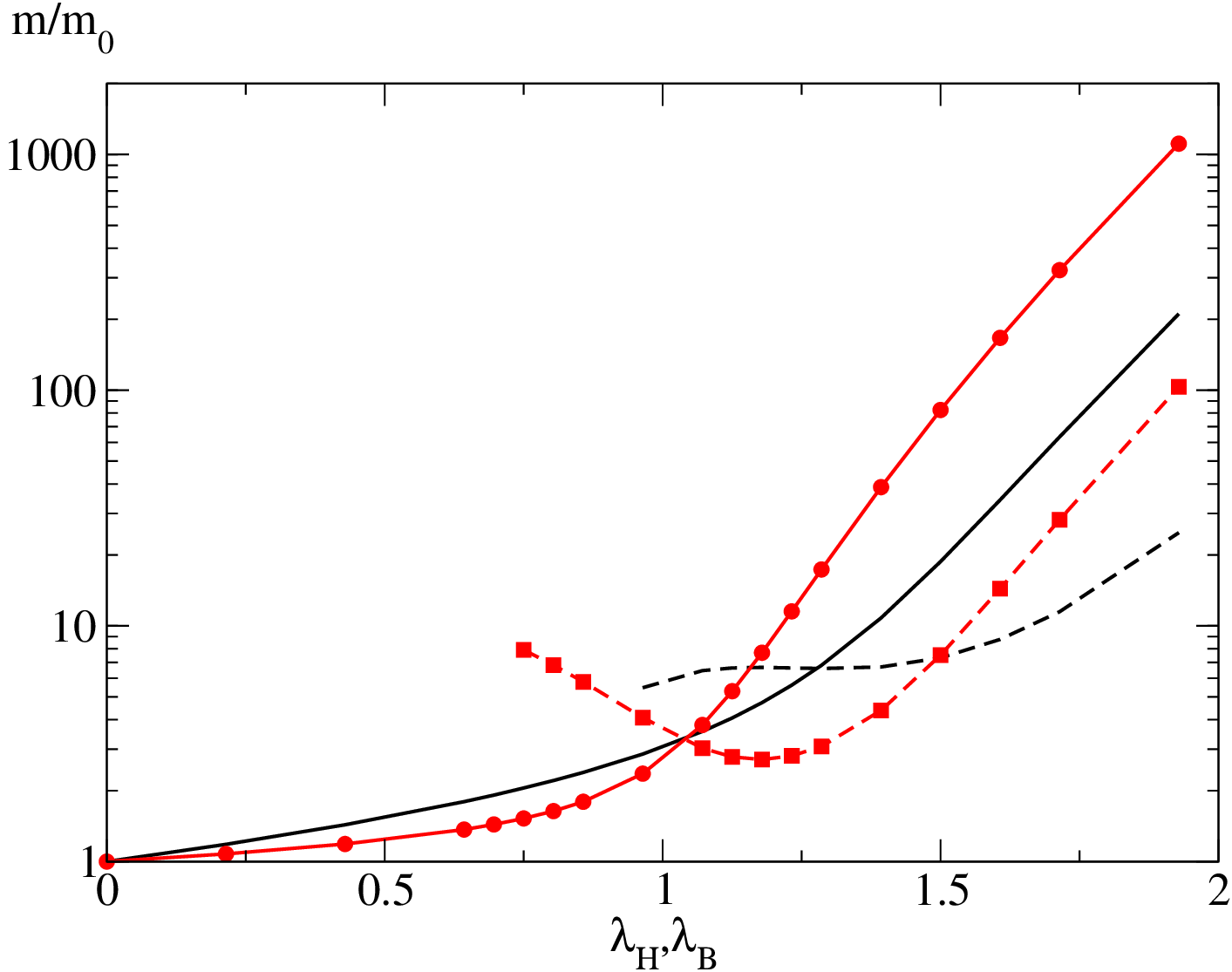}
\caption{\label{fig:mass} (color online) Ratio of effective polaron
  mass to that of the free electron. Circles and squares show
  breathing-mode results for GS and first bound state,
  respectively. The other lines correspond to the Holstein model GS
  (full) and second bound state (dashed). Parameters are $t=2,
  \Omega=1$.}
\end{figure}

The effective masses for the two lowest eigenstates of both models are
shown in Fig.~\ref{fig:mass}, as a function of $\lambda_{H,B}$.  These
were calculated from the second derivative of the energy at momentum
$K=0$. For the GS of both models, the effective mass increases
monotonically with $\lambda_{H,B}$. At weak couplings, the
breathing-mode polaron is lighter than the Holstein polaron. As
already discussed, this is due to the  vanishingly weak
coupling to low-momentum phonons. At strong coupling, however, the
effective mass is larger for the breathing-mode polaron. This is in
agreement with predictions of the strong-coupling perturbation theory,
and results from the fact that the hopping of a breathing-mode polaron
involves phonon clouds on $2z-1$ phonon sites, whereas hopping of a Holstein
polaron involves phonons at only two sites.

The effective mass of the first excited state can only be defined once
this state has split-off from the continuum. It has non-monotonic
behavior, first decreasing and then increasing with increasing
$\lambda_{H,B}$. This can be understood through the link of the
effective mass and the $qp$ weight. In terms of derivatives of the self-energy
$\Sigma(k,\omega)$,  the
effective mass $m^*$ is given by:
\begin{equation}
{m^* \over m} = \left(1- \frac{\partial \Sigma}{\partial
  \omega}\right) \cdot \left(1+ {m\over
  \hbar^2}\frac{\partial^2\Sigma}{\partial k^2}\right)^{-1},\nonumber
\end{equation}
where derivatives are evaluated at $K=0$ and at the corresponding
eigenenergy. The first term is linked to the $qp$ weight, $Z = (1-
\frac{\partial \Sigma}{\partial \omega})^{-1}$, so that $m^* \sim
1/Z$. As shown in Fig. \ref{fig:1st}(b), the $qp$ weight of the
first excited state has non-monotonic behavior, leading to the
non-monotonic behavior of the effective mass.

\begin{figure}[b]
\includegraphics[width=0.8\columnwidth]{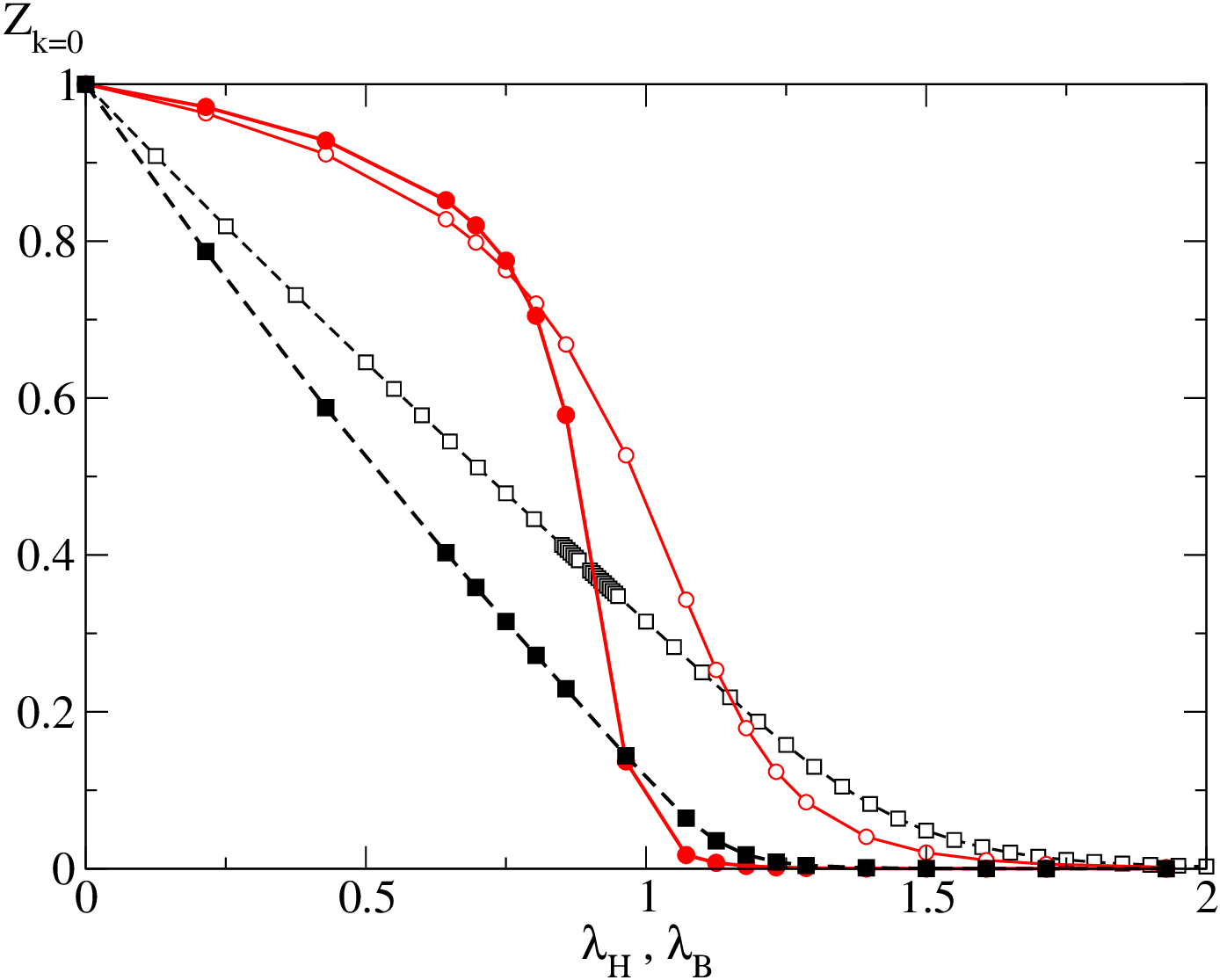}
\caption{\label{fig:compt} (color online) GS $qp$ weights for the
  breathing-mode (circles) and Holstein (squares) models. Full symbols
  correspond to $\alpha=8$. For comparison purposes, the empty symbols show
  the $\alpha=4$ results of Fig. \ref{fig:gs}(b). ($\Omega=1$)}
\end{figure}

All the results shown so far were for $\alpha=4$. For higher $\alpha$
(lower $\Omega$ and/or larger $t$),
the difference between the two models can be grasped from
Fig.~{\ref{fig:compt}}. Similar to the Holstein model, the
large-to-small polaron transition occurs at lower $\lambda$ for
increasing $\alpha$.\cite{v6,a2} At weak and moderate coupling, the $qp$ weight
and the effective mass (not shown) in
the breathing-mode model are much less sensitive to an increase in
$\alpha$, than is the case for the Holstein polaron. This suggests that
breathing-mode polarons should be better charge carriers than the Holstein
polarons, in this regime.

\subsection{\label{sec:sf}Spectral Function}

\begin{figure}[t]
\includegraphics[width=0.9\columnwidth]{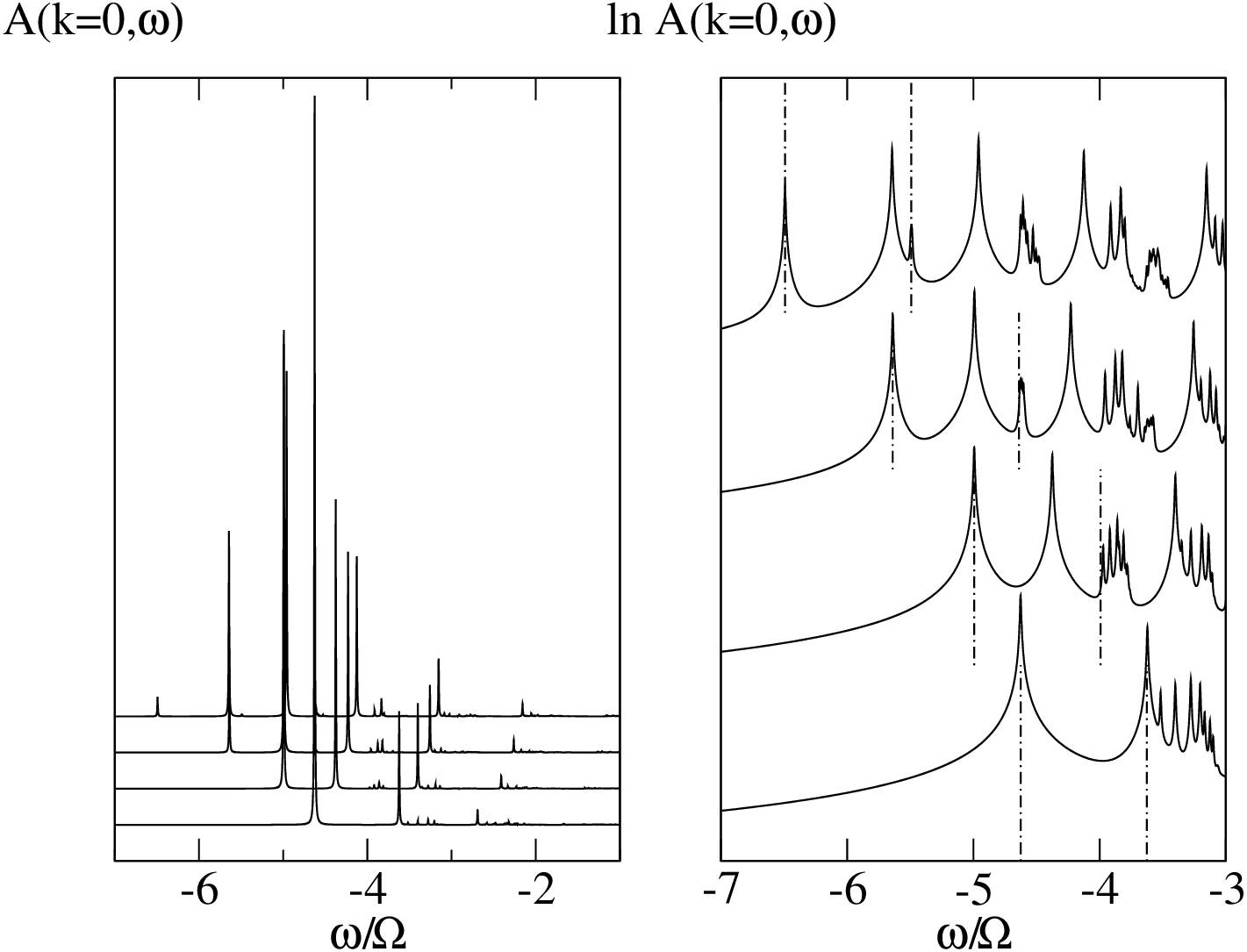}
\caption{\label{fig:abl} The spectral function of the 1D
breathing-mode model, on linear (left) and logarithmic (right)
scales. The curves are shifted for better viewing, and correspond
(top to bottom) to $\lambda_B=1.5; 1.25; 1.0$ and $0.75$. Vertical
lines indicates $E_{GS}$ and $E_{GS}+\Omega$. $K=0, t=2,
\Omega=1, \eta = 0.004\Omega$. }
\end{figure}

The spectral function is proportional to the imaginary part of the
Green's function:
\begin{equation}
A(k,\omega)=-\frac{1}{\pi}\mbox{Im}G(k,\omega)
\end{equation}
In terms of single electron eigenstates and eigenfunction $H|n\rangle
= E_n |n\rangle$, we obtain the Lehmann representation:
$$ A(k,\omega)=\sum_n |\langle n|c^{\dag}_k|0\rangle|^2\delta(\omega-E_n)
$$ Of course, since we use a finite small $\eta$ in numerical
calculations, the $\delta$-functions are replaced Lorentzians of width
$\eta$ [see Eq. (\ref{eq:gf})]. We calculate the Green's functions as
discussed in the previous section.

Fig.~\ref{fig:abl} shows the spectral function for zero momentum as a
function of energy. Results corresponding to four different coupling
strengths $\lambda_B$ from near the crossover region are shown for the
breathing-mode polaron.  We note that there is always a continuum
starting at one phonon quantum above the ground state energy. This is
more clearly visible in the right panel, where the spectral weight is
shown on a logarithmic scale, and vertical lines mark the position of
the ground-state energy $E_{GS}$, respectively of $ E_{GS}+\Omega$. As
the coupling $\lambda_B$ increases, we see the appearance of the
second bound state below the continuum. We only find at most one extra
bound state in this energy range for all coupling strengths. As
$\lambda_B$ increases, the spectral weight of the first continuum
decreases dramatically. Other bound states form above it, followed by
higher energy continua whose weight is also systematically
suppressed. This is qualitatively similar to the behavior exhibited by
Holstein polaron.\cite{a3}

\begin{figure}
\includegraphics{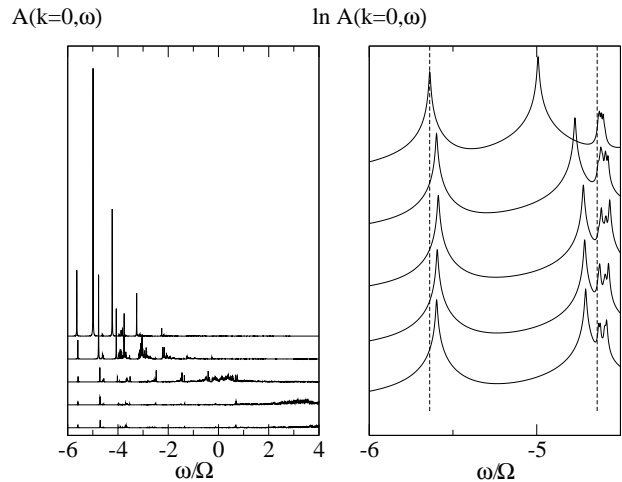}
\caption{\label{fig:abk} $A(k,\omega)$ vs $\omega$, for  $K/\pi$=0;
  0.25; 0.5; 0.75 and 1 (top to
bottom) for intermediate coupling $t=2, \Omega=1,
\lambda_B=1.25$. $\eta/4=\Delta E=0.001\Omega$. The height of the
spectral weight is plotted in linear scale on the left and logarithmic
scale on the right. The two vertical lines indicate $E_{GS}$ and
$E_{GS}+\Omega$.}
\end{figure}
Fig.~\ref{fig:abk} illustrates the momentum dependence of the
breathing-mode polaron's spectral weight. The results correspond to a
coupling above the critical value, where there is a second bound
state. The majority of the spectral weight is transferred to much
higher energies as the momentum increases, and a broad feature
develops at roughly the position of the free-electron energy for that
momentum. This spectral weight transfer is also qualitatively similar to
what is observed for Holstein polarons.  Our results have a
high-enough resolution to clearly show the continuum at
$E_{GS}+\Omega$ for all values of $K$.  This is part of the kink-like
structure reported in Ref. \onlinecite{scba}. The logarithmic plot
clearly reveals a non-monotonic dispersion of the ground-state like
in Fig.~\ref{fig:disp}, characteristic for the breathing-mode polaron.

Fig.~\ref{fig:abk} shows only one peak located between the
ground-state and the polaron+one-phonon continuum, even though in
fact we believe that there are more than one eigenstates within this
region. We found, from eigenvalue computation, additional energy states
below the continuum; however, computation of exact energy values requires
prohibitively long compution time due to the clustering of eigenvalues.
By observing the convergence behavior due to increasing basis size, we can
conclude that additional bound states do exist below the continuum.
The lack of their contribution to the spectral function can be understood by
the fact that the single particle Green's function only contains
information about eigenstates with finite $qp$ weight, $|\langle \phi | c_K^\dag|0\rangle|^2 > 0$, see Eq.~(\ref{eq:gf}). These eigenstates must have
components corresponding to some Lang-Firsov eigenstate with no off-site
phonons (Eq.~(\ref{eq:transform})). Also, the wave-function of these
states must have a peculiar space inversion symmetry: S-symmetric
for all even-phonon-number components and P-symmetric for all
odd-phonon-number components. The ground-state always satisfies this
requirement, but above a critical coupling, only one other state
below the phonon threshold satisfies this requirement.

\section{\label{sec:conc} Conclusions}

In summary, we have reported here the first accurate numerical study
of the 1D breathing-mode polaron. A previous study\cite{scba} based on
the self-consistent Born approximation proves to be inadequate to
describe correctly the behavior for medium and strong couplings, as
expected on general grounds.

Comparison with the Holstein model results, which correspond to a
coupling $g(q) = const.$, reveals some of the similarities and
differences of the two models. The breathing-mode polaron is much
more robust (has a much larger $qp$ weight, and less variation with
parameters) at weak couplings. This is a direct consequence of the
fact that coupling to low-momentum phonons, which is relevant here,
becomes vanishingly small $g(q) \sim \sin(q/2) \rightarrow 0$.
Similar behavior is expected for any other $g(q)$ model if
$\lim_{q\rightarrow0} g(q) \rightarrow 0$. On the other hand, at
strong couplings the breathing-mode polaron is much heavier and has
a lower $qp$ weight than the Holstein polaron. This also results
from strong-coupling perturbation results, and is due to the fact
that in order to move from site $i$ to site $i+1$, a small
breathing-mode polaron must (i) create a new polaron cloud at site
$i+\frac{3}{2}$; (ii) rearrange the polaron cloud at site
$i+{1\over2}$, so that its displacement is now pointing towards site
$i+1$, not towards site $i$; and (iii) relax (remove) the phonon
cloud at site $i-{1\over 2}$. This results in a suppressed polaron
kinetic energy, and an enhanced effective mass. Because of the
larger $Z$ at weak coupling, and the lower $Z$ at strong couplings,
the crossover from large to small polaron is much sharper for the
breathing-mode polaron. Another interesting observation is that the
polaron dispersion becomes non-monotonic with momentum $k$ for
medium and large couplings. This can be understood in terms of
strong-coupling perturbation theory, which shows that the
second-order $\cos(2k)$ correction is larger than the first-order
$\cos(k)$ correction for large-enough couplings.

Similarities with the Holstein behavior regard the appearance of the
polaron+free phonon continuum at $E_{GS}+\Omega$, and the appearance
of a second bound state with finite $qp$ weight for large-enough
couplings. The convergence of numerics points to the existence of additional
bound states, whose absence from the spectral function can be explained by symmetry or missing free-electron componenets in the wavefunction; however, this
issue is not fully settled. Also, the importance of such states for the physical
properties is not known. The general aspect of the higher energy spectral
weight at strong couplings, as a succession of bound states with
large spectral weight and continua with less and less spectral weight,
is also reminiscent of the Holstein polaron results.

{\bf Acknowledgments:} We thank Jeremy Heyl for giving us permission
to use his cluster. This work was supported by CFI (access to the
clusters yew, westgrid and jmd), by NSERC, and by CIfAR Nanoelectronics
and the Alfred P. Sloan Foundation (M.B.) and CIAR Quantum Materials
(G.S.).


\begin{references}
\bibitem{note1} We have reported a software bug in the parallel
version of the ARPACK package, but the package has not been updated
as of to date.
\end{references}
\end{document}